\documentclass[11pt,letterpaper,onecolumn]{article}
\usepackage{indentfirst}
\usepackage{url}
\usepackage[top=2cm,bottom=2cm]{geometry}
\usepackage{comment}
 
\newcommand{\p}{\partial}

\usepackage{answers}
\usepackage{authblk}
\usepackage{cancel}
\usepackage{setspace}
\usepackage{blindtext, rotating}
\usepackage{graphicx}
\usepackage{enumitem}
\usepackage{textcomp}
\usepackage{multicol}
\usepackage{mathrsfs}
\usepackage{amsmath,amsthm,amssymb}
\usepackage{indentfirst}
\usepackage{mdframed}
\usepackage{tikz-cd}
\usepackage{fancyhdr}
\usepackage{tensor}
\usepackage{braket}
\usepackage{float}
\usepackage[squaren]{SIunits}
\title{\bf On the relationship between the gain of Stimulated Raman  Backscatter  of a short-pulse laser in a plasma and the Pierce parameter for XFELs
}

\author[1,2]{Xinlu Xu}
\author[3,4]{Warren B. Mori}
\affil[1]{State Key Laboratory of Nuclear Physics and Technology, and Key Laboratory of HEDP of the Ministry of Education, CAPT, School of Physics, Peking University, Beijing 100871, China}
\affil[2]{Beijing Laser Acceleration Innovation Center, huairou, Beijing 100871, China}
\affil[3]{Department of Physics and Astronomy, University of California, Los Angeles, California 90095, USA}
\affil[4]{Department of Electrical and Computer Engineering, University of California, Los Angeles, California 90095, USA}

\begin{document}
\maketitle
%=====================================================
%============ Abstract ===============================
%=====================================================

\begin{abstract}
It is shown that the convective gain of Stimulated Raman Backscatter (SRBS) within a single laser wavelength in the strongly coupled regime is equivalent to the Pierce parameter commonly used to describe high gain x-ray Free Electron Lasers (XFELs). The comparison utilizes the fact that the gain is a Lorentz invariant so the Pierce parameter is written in terms of the FEL parameters in the electron beam's rest frame.  It is thus inferred that there is a direct correspondence of the physics between SRBS in the strong-coupled regime and high gain XFELs in the Compton regime. 
\end{abstract}

%=====================================================
%============ Body of the article ==========================
%=====================================================

%=====================================================
%============ Section ==================================
%=====================================================

\section{Introduction}
This article is based on notes for a lecture that one of the authors has given several times at a high energy density plasma summer school sponsored by the Center for Matter at Extreme Conditions. The motivation was to show that the physics involved in Stimulated Raman Backscatter (SRBS) of a short-pulse laser \cite{Short_pulse_SRS_Darrow} in a plasma is essentially identical to that of the Free Electron Laser (FEL) instability in the high gain Compton regime \cite{Kroll1978PhysRevA.17.300}. This regime of an FEL is the basis for all FELs that generate coherent x-rays (XFELs) operating worldwide. Furthermore, this relationship has also recently become of interest because the same physics arises when an electromagnetic wave backscatters in an electron-positron (pair) plasma. Although this relationship has been known since the early 1980s, the authors are not sure where it was first discussed in the literature. The purpose of this note is not only to comment on this relationship, but also to show that, if the proper regimes of growth are identified, using SRBS theory in a plasma gives the exact same number of e-foldings as that using FEL theory in its relevant regime. 

There is an extensive literature of both Stimulated Raman Scattering (SRS) \cite{forslund1975theory, kruer1988physics, michel2023introduction} and references therein, and FELs \cite{Kroll1978PhysRevA.17.300, pellegrini2016physics, kim2017synchrotron} that spans nearly half a century. The purpose of this note is not to review this work nor to be comprehensive and cover all regimes of SRBS and FELs, but rather to emphasize the regimes of current relevance. At first glance, it may seem surprising that there is any connection in the physics of SRBS and an XFEL. In SRBS an incoming electromagnetic wave (EM), pump laser,  scatters off of a forward propagating (with respect to the incoming laser)  electron plasma wave (EPW) into another EM wave. It is called backscatter (forward scatter) if the scattered EM wave propagates backward (forward). Fundamental to this instability is frequency and wavenumber matching between the three-waves, i.e., $\omega_0=\omega_1+\omega_2$ and $k_0=k_1+k_2$, where the subscripts $0,1,2$ correspond to the pump, backcattered, and electron plasma waves (EPW) respectively. The frequency and wavenumbers of each wave nearly satisfy the appropriate dispersion relation in the plasma. If frequency and wave number matching exists then the beating of the two light waves produces a radiation pressure (ponderomotive) force that resonantly drives the EPW. The beating of the pump and the EPW then resonantly increases the scattered EM wave. This feedback is the basis for the SRBS instability which has been studied extensively primarily due to its implications for inertial fusion energy (IFE) research. 

On the other hand, in the FEL instability a relativistic electron (e-) beam propagates through an undulator. A typical or conventional undulator is a static magnetic field oriented perpendicular to the propagation direction of the e-beam with a well defined wavenumber in the propagation direction. As the e-beam oscillates transversely (wiggles) it radiates in the forward direction. The wavelength of the radiation will match the distance the e-beam slips backwards during the time it takes to move through one undulator period [$2\pi/(k_uv_b)$] leading to the relationship $\lambda_r\approx \lambda_u/2\gamma_b^2$.  The beating of the motion in the undulator and the coherent x-ray radiation provides a ponderomotive force which bunches the e-beam so that it coherently radiates. The resonance condition is $(k_r+k_u)v_b=\omega_r$ which also gives $\lambda_r\approx \lambda_u/2\gamma_b^2$.  For now we assume that the normalized momentum oscillation amplitude in the undulator, $K_0\equiv p_u/mc=eB/k_umc^2$, is $\ll 1$. In reality, our assumption holds if $K_0\lesssim 0.5$. The undulator can also be a counterpropagating laser \cite{elias1979high, gover1982feasibility, PhysRevAccelBeams.27.011301}.

A hint for the connection between SRBS and the FEL instability can be seen if one views the FEL process in the frame of the e-beam. In this case the undulator transforms to a quasi-EM wave (the dispersion relation is $\omega'=v_b k'$ and not $\omega'=ck'$) that propagates towards the beam. The resulting radiation has nearly the same frequency as that of the Lorentz transformed undulator, but it propagates backwards in this frame. Transforming this radiation back into the lab frame provides the $\gamma_b^2$ scaling for the upshift. 

In this note, we demonstrate that the physics of these two distinct processes is the same by first identifying the relevant regimes for SRBS and the XFEL. SRBS can occur in a variety of regimes based on whether the laser pulse is long or short compared to the plasma length, whether convective or absolute growth is occurring, or if the backscattered radiation is growing temporally, spatially, or spatial-temporally. 

To determine if the long or short pulse regime of SRBS is of interest, we first examine if the undulator is long or short compared to the e-beam length in the relevant frame. Consider the parameters of interest at the LCLS \cite{Emma2010NP}. In the lab frame at the LCLS the electron beam has an energy of 10 GeV or a relativistic factor of $\gamma_b\approx 2\times 10^4$,  and a bunch length of $\sigma_z\approx30~\micro\meter$ with $\approx 1.5\times10^9$ electrons,  where the density profile is of the form $n_b=n_0e^{-z^2/2\sigma_z^2}$ and $n_0=N/[(2\pi)^{3/2}\sigma_z\sigma_r^2]$.  The undulator has a length, $L=N\lambda_u\approx 100~\meter$ where $\lambda_u=3~\centi\meter$ and $N$ is the number of undulator periods. Thus, in the lab frame $\sigma_z \ll L$. However, in the frame of the beam, the length of  the undulator is Lorentz contracted to $L'=L/\gamma_b=0.5~\centi\meter$ while the length of the beam is Lorentz expanded to $\sigma_z'=\gamma_b\sigma_z=20~\centi\meter$.  Thus, in the beam's frame we are in the opposite limit  where $L' \ll \sigma_z'$ or the undulator looks like a short pulse laser moving towards a relatively long electron beam. Although not often described in this manner, these are the conditions to be in FEL regime where the Pierce parameter \cite{rhopara1984, pellegrini2016physics} (will be carefully defined below) is the figure of merit for the gain. 

With this insight, we start from the coupled equations for SRBS and then use these to obtain envelope equations for SRBS in the strongly coupled regime. This regime is generally of relevance when the laser frequency is much greater than the EPW frequency, $\omega_0\gg \omega_p$ where $\omega_p^2\equiv 4\pi e^2n_0/m$ is the plasma electron frequency in the rest frame and $n_0$ is the plasma density. For the XFEL parameters, in the frame of the beam $\sigma_r'=\sigma_r \approx 10~\micro\meter$ and $\omega_b'=\omega_b/\sqrt{\gamma_b}\approx 7\times10^{10}~\radian\per\second$ where $\omega_u'\approx1.3\times10^{15}~\radian\per\second$, and thus $\omega_u'\gg \omega_b'$.

We then make a mathematical transformation into speed of light variables that measure the location within the laser in the lab frame with respect to the head of the laser and measure the distance the laser has propagated into the plasma. This co-moving distance from the head of the laser is the length that determines the number of e-foldings across a short-pulse laser. Utilizing the fact that the gain is a Lorentz invariant, we next rewrite the Pierce parameter using the beam and undulator parameters in the rest frame of the beam and show that after carefully accounting for numerical factors this expression is identical to the expression for the gain of short-pulse SRBS in the strongly coupled regime. Last, we comment on differences between the instabilities when other realities are considered and the implications of these results for SRBS in pair plasmas.

\section{Equations of SRBS} 
We start from the coupled equations based on fluid theory for SRBS \cite{forslund1975theory, kruer1988physics, michel2023introduction},
\begin{align}
    (\p_t^2 - c^2 \nabla^2 + \omega_{p}^2)\mathbf{a}_0 &= -\omega_{p}^2 n_e \mathbf{a}_1, \\
   (\p_t^2 - c^2 \nabla^2 + \omega_{p}^2)\mathbf{a}_1 &= -\omega_{p}^2 n_e^* \mathbf{a}_0, \\
   (\p_t^2 - \gamma_e v_e^2 \nabla^2 + \omega_{p}^2)n_e &= c^2 \nabla^2(\mathbf{a_0}\cdot \mathbf{a}_1),
\end{align}
where $\mathbf{a}_j\equiv e\mathbf{A}_j/mc^2$ for $j=0, 1$ is the normalized vector potential of an EM wave, $n_e$ is the density perturbation of the electrons, $v_e$ is the plasma electron thermal velocity, and $\gamma_e$ is the electron adiabatic index, with $\gamma_e=3$ for one-dimensional (1D) electron thermal motion. The term that drives $n_e$ is the divergence of the time averaged ponderomotive force or radiation pressure, $F_p=mc^2\nabla \langle\mathbf{a_0}\cdot\mathbf{a_1}\rangle$ where $\langle \rangle$ defines averaging over the pump oscillations. We assume linearly polarized light but it is trivial to extend this to circularly polarized pumps. Deriving these equations also relies on the fact that the transverse canonical momentum of electrons is approximately conserved (it is perfectly conserved in 1D) from which it follows that electrons wiggle in the laser field as $p_j/mc=eA_j/mc^2\equiv a_j$.

Next, we assume the fields are of the one dimensional form, 
\begin{align}
    \mathbf{a}_0 &=\hat x \frac{b_0(z,t)}{2}\exp[i(\omega_0t - k_0z)] + \mathrm{c.c.},\\
    \mathbf{a}_1 &=\hat x \frac{b_1(z,t)}{2}\exp[i(\omega_1t + k_1z)] + \mathrm{c.c.}, \\ 
    n_e &=\frac{b_2(z, t)}{2}\exp[i(\omega_2 t - k_2 z)] + \mathrm{c.c.},
\end{align}
and where we have let the wave amplitudes be general functions of $z$ and $t$. It is also assumed that frequency and wavenumber matching hold (as was described above). 

There are two ways to proceed. The first is to let the amplitudes be constant by letting $\omega_j$ and $k_j$ be complex. The second is to let the amplitudes vary in $z$ and $t$ and assume that $\omega_j$ and $k_j$  are real and obey the linear dispersion relations.  We adopt the second approach. We also assume the variation of the amplitudes in time and space are smaller than that of the frequencies and wavenumbers for both light waves, $|\p_tb_j|\ll|\omega_jb_j|$ and $|\p_zb_j|\ll|k_jb_j|$ for $j=0$ and 1. This is called the envelope approximation. We do not make this assumption for the EPW. We are thus left with the following coupled equations,
\begin{align}
    2i\omega_0(\p_t + v_{g0}\p_z)b_0 &=-\frac{\omega_p^2}{2}b_2b_1, \\
    2i\omega_1(\p_t - v_{g1}\p_z)b_1 &=-\frac{\omega_p^2}{2}b_0b_2^*, \\
    [\p_t^2-3v_{e}^2\p_z^2+2i\omega_2(\p_t+v_{g2}\p_z)]b_2 &= -\frac{c^2k_2^2}{2}b_0b_1^*,
\end{align}
where $v_{gj}$ is the magnitude of the group velocity of mode $j$. For simplicity we will henceforth assume that the plasma is cold, $v_{e}\rightarrow 0$ and that $v_{gj}\rightarrow c$ for $j=0$ and 1. We neglect pump depletion when determining the gain, resulting in the two coupled equations,
\begin{align}
    2i\omega_1 (\p_t - c\p_z) b_1 &=-\frac{\omega_p^2}{2}b_0b_2^*, \label{eq:b1} \\
    (\p_t^2 + 2i\omega_2\p_t+\nu)b_2 &= -\frac{c^2k_2^2}{2}b_0b_1^* \label{eq:b2},
\end{align}
where we added a phenomenological damping term $(\nu)$ to remove some pathological behavior that can arise in the cold plasma limit for spatial growth. These equations describe various regimes of growth: 1) Temporal growth where $\p_z\rightarrow 0$, 2) Spatial growth where  $\p_t\rightarrow 0$, and 3) Spatial-temporal growth.

\begin{figure}[htbp]
\centering
\includegraphics[width=0.9\linewidth]{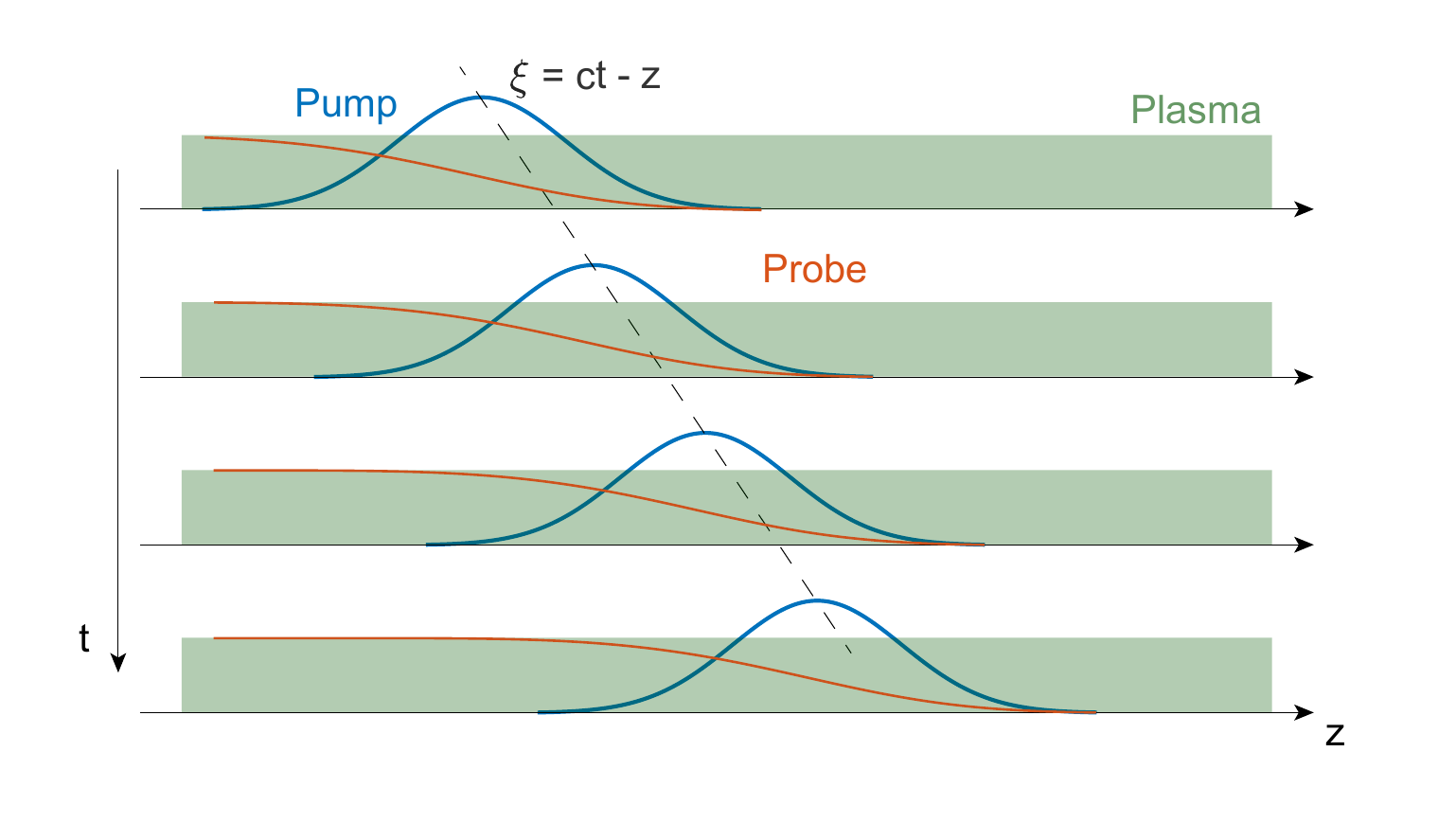}
\caption{\label{fig:SRBS} Schematic of SRBS in the short-pulse limit where the gain is a function of $\xi$. In the short-pulse regime the backscattered wave grows from the front to the back of the pump and the evolution is static when viewed in the $\xi=ct-z$ variable.}
\end{figure}

We are interested in the short-pulse regime, where spatial-temporal growth occurs. In this regime a laser keeps propagating into unperturbed (fresh) plasma. The growth arises from a noise source at the head of the pump laser. This source can be random or a small well defined coherent source. The backscattered light then convects backwards from the head of the laser and gets amplified until it leaves the rear of the laser. Thus, the picture looks static with respect to the variable $\xi\equiv ct-z$. If we transform from the variables $z,t$ to $\xi\equiv ct-z, s\equiv z$ 
then the growth only occurs in $\xi$ and the picture does not change (it is static) as the laser propagates deeper into the plasma. This is illustrated in Fig. \ref{fig:SRBS}. We thus make a mathematical transformation into the speed of light variables using the relationships, $\p_t=c\p_{\xi}$ and 
$\p_z=-\p_\xi + \p_s$, leading to,
 \begin{align}
    4ic\omega_1 \p_{\xi} b_1 &= - \frac{\omega_p^2}{2}b_0b_2^*, \\
    (c^2\p_{\xi}^2 + 2ic\omega_2\p_{\xi}+\nu)b_2 &= -\frac{c^2k_2^2}{2}b_0b_1^* .
\end{align}
In this frame there are only partial derivatives with respect to $\xi$, i.e., $\p_{\xi}$, so they can be replaced by total and not partial derivatives, $\mathrm{d}_{\xi}$. 

We next consider two regimes using the nomenclature from plasma physics; these are the weakly and strongly coupled regimes; and the damping term is dropped, $\nu\rightarrow0$. In the weakly coupled regime $\mathrm{d}_{\xi}\ll \omega_p$ so that we drop the $\mathrm{d}_{\xi}^2$ term for $b_2$ to obtain,
\begin{align}
    4ic\omega_1 \mathrm{d}_{\xi} b_1 &= - \frac{\omega_p^2}{2}b_0b_2^*, \\
    2ic\omega_2 \mathrm{d}_{\xi}b_2 &= - \frac{c^2k_2^2}{2}b_0b_1^*,
\end{align}
and these can be combined into a single equation for $b_1$ (or $b_2$),
\begin{align}
    (c^2\mathrm{d}_{\xi}^2 - \gamma_0^2/2)b_1=0,
\end{align}
where $\gamma_0\equiv \frac{\omega_p ck_2}{4\sqrt{\omega_1\omega_2}}|b_0|$ is the temporal growth rate [recovered from Eqs. \eqref{eq:b1} and \eqref{eq:b2} when $\p_z$ terms are dropped]. For cold plasmas and low densities $\omega_2=\omega_p, k_2\approx 2 k_0$ and $ck_1\approx ck_0\approx \omega_0$ such that $\gamma_0\approx \sqrt{\omega_0\omega_p} |b_0|/2$.  

The spatial-temporal gain is reduced by a factor of $2^{1/2}$ from the temporal growth times the pulse length. This reduction arises because the laser and backscattered radiation are moving towards each other. The gain of SRBS in the short-pulse limit can be written as $b_1=b_1 (0) \exp\left( 2^{-1/2}\gamma_0\sigma_z/c \right)$, where $\sigma_z$ is the pulse length and $b_1(0)$ is the noise at the head of the laser. We can use this expression to determine the conditions for which the weakly coupled assumption breaks down. Computing the ratio of $\mathrm{d}_{\xi}b_1(\xi)$ to $\omega_pb_1$, we find the condition to be 
%for being 
in the weakly coupled regime,
\begin{align}
    (\omega_0/\omega_p)^{1/2}|b_0| \ll 1.
\end{align}

However, for very low density plasmas and/or sufficiently high pump strengths (which is the corresponding case for XFELs) this condition is not satisfied and SRBS is in the strongly coupled limit where $|\mathrm{d}_{\xi}| \gg \omega_2$. In this regime the coupled equations are instead,
\begin{align}
    4ic\omega_1 \mathrm{d}_{\xi}b_1 &= -\frac{\omega_p^2}{2}b_0b_2^*, \\
    c^2 \mathrm{d}_{\xi}^2b_2 &= -\frac{c^2k_2^2}{2}b_0b_1^*, 
\end{align}
which can be combined into,
\begin{align}
  \left( c^3 \mathrm{d}_{\xi}^3 + i\frac{1}{4}\omega_p^2\omega_0 |b_0|^2\right) b_1=0,
\end{align}
where we have made the cold plasma, low density approximations. There are three solutions to this equation and the one that corresponds to growth as the scattered light convects backwards is,
\begin{align}
  b_1=b_1(\xi=0)\exp\left[\left(\frac{\sqrt{3}}{2} - \frac{i}{2}\right)\left(\frac{|b_0|^2}{4}\omega_p^2\omega_0\right)^{1/3}\frac{\xi}{c}+(i\omega_1 t + i k_1 z)\right],
\end{align}
where the number of e-foldings is given by,
\begin{align}
  \frac{\sqrt{3}}{2}\left(\frac{1}{4}|b_0|^2\omega_p^2\omega_0\right)^{1/3}\sigma_z/c. \label{eq:SRBS_G}
\end{align}

Armed with Eq. \eqref{eq:SRBS_G} we can now proceed to relate the SRBS gain in the strongly coupled short-pulse regime to the Pierce parameter.

\section{Comparison of the Gain for SRBS to the Pierce parameter}
Our conjecture is that the gain of an XFEL in the frame of the beam can be exactly mapped to that of SRBS in the relevant regime. It is therefore important to be precise when defining the gain and relating it to the Pierce parameter. In the beam's frame, the undulator is either a quasi-EM or pure EM mode propagating at a stationary electron beam. In this frame all quantities are denoted with primes, i.e., $'$s.  The effective length of the pump laser for an XFEL is $N\lambda_u'$ where $N$ is the number of wavelengths of the undulator which is a Lorentz invariant and $\lambda_u'=\lambda_u/\gamma_b$ is the Lorentz transformed undulator wavelength. The undulator strength $K_0$ or $b_0$ (normalized vector potential transverse to the beam's velocity) is also a Lorentz invariant. To facilitate comparison to the Pierce parameter we use wave numbers instead of frequencies. The wave number of the pump transforms as $k_u'=\gamma_b k_u$ and the plasma wavenumber, defined as $k_p\equiv \omega_p/c$,  transforms as $k_p'^2=k_p^2/\gamma_b$ where the proper density, $n/\gamma$, is another Lorentz invariant. We define the gain factor, $G$, as the number of e-foldings of growth for the field amplitude, $b_j \propto  \exp(G)$.   In the frame of the electron beam we can use Eqn. \eqref{eq:SRBS_G},
\begin{align}
  G'=\frac{\sqrt{3}}{2}\left(\frac{1}{4}|b_0|^2 k_p'^2 k_u' \right)^{1/3}N\lambda_u' = \frac{\sqrt{3}}{2}\left(\frac{1}{16}|b_0|^2\frac{k_p'^2 k'^2}{k_u'}\right)^{1/3}N\lambda_u' .
\end{align}\label{eq:Gprime}

It is also important to recognize that $G$ is a Lorentz invariant, i.e.,  $G'=G$, so we can obtain $G$ in terms of unprimed quantities by starting from the expression for $G'$ and then rewriting the $'$ quantities in terms of their unprimed quantities. When describing FELs it is common to use the gain for the power or intensity, and not for the wave amplitude, as the figure of merit. For historical reasons, the Pierce parameter is defined as the power gain over a distance $\frac{\lambda_u}{4\pi\sqrt{3}}$,
\begin{align}
  \rho\equiv \frac{2G}{N}\frac{1}{4\pi\sqrt{3}},
\end{align}
and upon substituting the expression for $G$ from Eqn. \eqref{eq:Gprime} we obtain,
\begin{align}
  \rho =\frac{1}{4\pi} \left(\frac{1}{4}|b_0|^2\frac{k_uk_p^2}{\gamma_b^3}\right)^{1/3}\lambda_u,
\end{align}
which upon some algebraic manipulation gives,
\begin{align}
\rho=\left(\frac{1}{32}|b_0|^2\frac{k_p^2}{k_u^2}\frac{1}{\gamma_b^3}\right)^{1/3}.
\end{align}

The Pierce parameter is often written in terms of the current and spot size of the beam. If the beam density in the lab frame is,
\begin{align}
n_b=\frac{N}{(2\pi)^{3/2}}\frac{1}{\sigma_r^2\sigma_z}e^{-\frac{r^2}{2\sigma_r^2}}e^{\frac{-\xi^2}{2\sigma_z^2}},
\end{align}
then after more algebraic manipulation, we obtain,
\begin{align}
\rho=\left(\frac{1}{32\pi}|b_0|^2\frac{I}{I_A}\frac{1}{2\pi\sigma_r^2}\frac{\lambda_u^2}{\gamma_b^3}\right)^{1/3},
\end{align}
where $I\equiv \frac{eN}{\sqrt{2\pi}\sigma_z/c}$ is the current of the beam and $I_A\equiv \frac{mc^3}{e}$ is the Alfven current.

This expression for the Pierce parameter is precisely the same as that obtained using the standard FEL treatments found in the more recent literature \cite{kim2017synchrotron}. We thus contend that the physics that drives current XFELs is precisely the same as that of strongly coupled short-pulse SRBS when viewed in the frame of the electron beam. In both cases the space charge forces that occur when the plasma or e-beam are modulated (bunched) do not affect the physics.

\section{Summary and closing comments}
In this note we have shown that the gain of SRBS of a short-pulse laser in the strongly coupled regime is exactly the same as that of an XFEL operating in the what is referred to as the Compton limit.\footnote{We note that the SRBS theory also has a Compton regime that has no connection to the XFEL regime with the same name; in SRBS Compton refers to  weakly coupled regime where the plasma has a relatively large temperature.}  It is thus argued that the fundamental physics is also the same. In this XFEL regime the gain can be parameterized in terms of the Pierce parameter which is defined as the power gain over a distance 
%every 
$\frac{\lambda_u}{4\pi\sqrt{3}}$. This demonstrates that the same physics is in play in both strongly coupled SRBS and the FEL process in the Compton regime. When viewed in the plasma or e-beam rest frame the growth rate of the instability is much larger than the natural oscillation frequency of the plasma or e-beam. Thus, the plasma or e-beam response is due to the beating of the pump and back scattered radiation, and not from its natural oscillations. 

This analysis is strictly valid in one-dimension, for cold plasmas, and for weak to moderate pump strengths. These approximations are generally reasonable for SRBS of a short pulse laser in the relevant regime. However, for XFELs the normalized pump strength is generally much larger than unity. In this case an analysis that includes relativistic corrections to the motion of beam electrons in the pump are needed. For example, in the lab frame the electrons in the beam will be slowed down as they enter the undulator (the energy remains constant but some is converted from axial to perpendicular motion) or if viewed in the frame of the electron beam, the incoming laser/undulator will cause the electron beam to drift forward with respect to the laser/undulator at a speed, $v_\mathrm{drift}=\frac{b_0^2/2}{(1+b_0^2/2)}$. This correction leads to the output radiation wavelength scaling as $(1+|b_0|^2/2)^{-1}$. The corrections to the Pierce parameter from both large pump strengths and multi-dimensional effects are known \cite{pellegrini2016physics}.  Esarey and Sprangle \cite{Esarey_harmonicsSRS_PRA} provided a nonlinear analysis of linearly polarized intense lasers interacting with a plasma or an electron beam. The analysis was motivated by harmonic generation of the backscattered radiation  but they also provided growth rates for nonlinear pump strengths. They did not comment on the invariance of growth rates when viewed in the rest frame of the plasma or electron beam and that the short pulse limit would be appropriate. 

An important difference between SRBS and an XFEL for large pump strengths is that while the undulator/laser can impart a drift to the e-beam associated with the incoming momentum of the pump this does not occur in a plasma because space charge from the more massive ions pulls the electrons back. The presence of the ions also prevents the electrons from losing energy which is why SRBS in a plasma can maintain resonance without the need of tapering the laser. 

Last, we comment on the implications of this work to SRS in a pair plasma. Recently, there has been renewed interest in how EM waves interact with pair plasmas because of Fast Radio Bursts (FRB) near magnetars \cite{Andersen2020, Mereghetti_2020, Bochenek2020}. If the FRB backscatters then the terrestial observations will be severely modified, thus backscatter of EM waves in pair plasmas is being investigated \cite{lyutikov2026, Nishiura}. In these astrophysical environments the plasma can be strongly magnetized ($\omega_p/\omega_c \ll 1$, where $\omega_c$ is the electron cyclotron frequency) and the normalized pump strength can be very large ($a_0 \gg 1$). However, as a starting point it is still useful to understand how an EM wave with moderate intensity interacts in an unmagnetized plasma. This is tantamount to seeing if SRS or stimulated Brillouin scattering (SBS) can occur in a pair plasma. In SBS the EM wave decays into a backscattered EM wave and a forward moving ion acoustic wave (IAW). In an EPW the electrons oscillate about the ions while in an IAW the electrons and ions oscillate together (even in an IWA the electrons and ions oscillate due to space charge forces). At first glance, it might seem that both modes would be supported in a pair plasma where in one case the electrons and positrons oscillate out of phase while in the  other they oscillate in phase. However, a kinetic analysis where the electrons and positrons have the same temperature shows that only the mode in which they oscillate out of phase exists. However, the ponderomotive force is identical (same magnitude and direction) on electrons and positrons so it cannot excite a mode where they oscillate out of phase.  Therefore, one might conclude that neither SRS nor SBS can occur. However, as we have shown above there is a strongly coupled regime of SRS (and also for SBS) where the pump strength dominates over any restoring forces of the natural oscillations. Thus,  this instability can still grow in a pair plasma where plasma bunching does not lead to any space charge restoring force. The only difference is that $\omega_p$ is replaced with $2^{1/2}\omega_p$ in the gain factor.  It can be argued that this is the pair plasma limit of either strongly coupled SRS or SBS,  as these instabilities merge together in a pair plasma in the strongly coupled regime. 

We also note that because in the XFEL instability the space charge force of the electron beam can also be ignored, we sometimes simulate the XFEL instability using a pair beam in one dimension where the density is appropriately  reduced by a factor of two to get the proper Pierce parameter. Recognizing the equivalence between the gain of the XFEL (related to the Pierce parameter) and SRS/SBS in a pair plasma allows one to take advantage of the modifications to FEL theory for large values of $b_0$. Additionally, in the pair plasma a large pump will cause the plasma to drift forward thereby doppler downshifting the reflected light in the lab frame. We and coauthors leave a more detailed analysis of SRS/SBS in a pair plasma for a future publication.

\section*{Acknowledgment}
We acknowledge useful conversations with life long colleagues including John M. Dawson, Chan Joshi, and Tom Katsouleas, recent conservations with Paulo Alves, Thomas Grismayer, and Vijay Patel,  and decades of support from the Department of Energy office of High Energy Physics and the National Science Foundation. 

\bibliography{refs}% Produces the bibliography via BibTeX.

\begin{thebibliography}{10}

\bibitem{Short_pulse_SRS_Darrow}
Chris Darrow, C~Coverdale, Michael Perry, W.~Mori, Chloe Clayton, K~Marsh, and
  Chandrashekhar Joshi.
\newblock Strongly coupled stimulated raman backscatter from subpicosecond
  laser-plasma interactions.
\newblock {\em Physical review letters}, 69:442--445, 08 1992.

\bibitem{Kroll1978PhysRevA.17.300}
Norman~M. Kroll and Wayne~A. McMullin.
\newblock Stimulated emission from relativistic electrons passing through a
  spatially periodic transverse magnetic field.
\newblock {\em Phys. Rev. A}, 17:300--308, Jan 1978.

\bibitem{forslund1975theory}
DW~Forslund, JM~Kindel, and EL~Lindman.
\newblock Theory of stimulated scattering processes in laser-irradiated
  plasmas.
\newblock {\em The Physics of Fluids}, 18(8):1002--1016, 1975; and references
  therein.

\bibitem{kruer1988physics}
W.L. Kruer.
\newblock {\em The Physics Of Laser Plasma Interactions}.
\newblock Avalon Publishing, 1988.

\bibitem{michel2023introduction}
Pierre Michel.
\newblock {\em Introduction to Laser-Plasma Interactions}.
\newblock Graduate Texts in Physics. Springer International Publishing, 2023.

\bibitem{pellegrini2016physics}
C.~Pellegrini, A.~Marinelli, and S.~Reiche.
\newblock {The physics of x-ray free-electron lasers}.
\newblock {\em Rev. Mod. Phys.}, 88:015006, Mar 2016.

\bibitem{kim2017synchrotron}
Kwang-Je Kim, Zhirong Huang, and Ryan Lindberg.
\newblock {\em Synchrotron radiation and free-electron lasers}.
\newblock Cambridge university press, 2017.

\bibitem{elias1979high}
Luis~R Elias.
\newblock High-power, cw, efficient, tunable (uv through ir) free-electron
  laser using low-energy electron beams.
\newblock {\em Physical Review Letters}, 42(15):977, 1979.

\bibitem{gover1982feasibility}
A~Gover, CM~Tang, and P~Sprangle.
\newblock Feasibility of dc to visible high-power conversion employing a
  stimulated compton free electron laser with a waveguided co2 laser pump wave
  and an axial electric field.
\newblock {\em Journal of Applied Physics}, 53(1):124--129, 1982.

\bibitem{PhysRevAccelBeams.27.011301}
Xinlu Xu, Jiaxin Liu, Thamine Dalichaouch, Frank~S. Tsung, Zhen Zhang, Zhirong
  Huang, Mark~J. Hogan, Xueqing Yan, Chan Joshi, and Warren~B. Mori.
\newblock Attosecond x-ray free-electron lasers utilizing an optical undulator
  in a self-selection regime.
\newblock {\em Phys. Rev. Accel. Beams}, 27:011301, Jan 2024.

\bibitem{Emma2010NP}
P.~Emma, R.~Akre, J.~Arthur, R.~Bionta, C.~Bostedt, J.~Bozek, A.~Brachmann,
  P.~Bucksbaum, R.~Coffee, F.-J. Decker, et~al.
\newblock First lasing and operation of an {\r a}ngstrom-wavelength
  free-electron laser.
\newblock {\em Nature Photonics}, 4(9):641--647, 2010.

\bibitem{rhopara1984}
R.~Bonifacio, C.~Pellegrini, and L.~M. Narducci.
\newblock Collective instabilities and high‐gain regime free electron laser.
\newblock {\em AIP Conference Proceedings}, 118(1):236--259, 1984.

\bibitem{Esarey_harmonicsSRS_PRA}
Eric Esarey and Phillip Sprangle.
\newblock Generation of stimulated backscattered harmonic radiation from
  intense-laser interactions with beams and plasmas.
\newblock {\em Phys. Rev. A}, 45:5872--5882, Apr 1992.

\bibitem{Andersen2020}
B.~C. Andersen, K.~M. Bandura, M.~Bhardwaj, A.~Bij, M.~M. Boyce, P.~J. Boyle,
  C.~Brar, T.~Cassanelli, P.~Chawla, T.~Chen, et~al.
\newblock A bright millisecond-duration radio burst from a galactic magnetar.
\newblock {\em Nature}, 587(7832):54--58, 2020.

\bibitem{Mereghetti_2020}
S.~Mereghetti, V.~Savchenko, C.~Ferrigno, D.~G{\"o}tz, M.~Rigoselli, A.~Tiengo,
  A.~Bazzano, E.~Bozzo, A.~Coleiro, T.~J.-L. Courvoisier, et~al.
\newblock Integral discovery of a burst with associated radio emission from the
  magnetar sgr 1935+2154.
\newblock {\em The Astrophysical Journal Letters}, 898(2):L29, jul 2020.

\bibitem{Bochenek2020}
C.~D. Bochenek, V.~Ravi, K.~V. Belov, G.~Hallinan, J.~Kocz, S.~R. Kulkarni, and
  D.~L. McKenna.
\newblock A fast radio burst associated with a galactic magnetar.
\newblock {\em Nature}, 587(7832):59--62, 2020.

\bibitem{lyutikov2026}
Maxim Lyutikov and Victor Gurarie.
\newblock Anderson self-localization of light in pair plasmas.
\newblock 2026.

\bibitem{Nishiura}
Rei Nishiura, Shoma~F. Kamijima, and Kunihito Ioka.
\newblock Unified kinetic theory of induced scattering: Compton, brillouin, and
  raman processes in magnetized electron and positron pair plasma.
\newblock {\em Phys. Rev. D}, 113:123070, Jun 2026.

\end{thebibliography}

\end{document}